\documentclass[twocolumn,3p,times]{elsarticle}
  
\usepackage{mathptmx, courier, pifont}
\usepackage[scaled=0.92]{helvet}
\usepackage[T1]{fontenc}
\usepackage{textcomp}
\usepackage{amsfonts}

\usepackage{epsfig}
\usepackage{amssymb}
\usepackage{graphicx}
\usepackage{amssymb}
\usepackage{graphicx}
\usepackage{mathtools, cuted}
\biboptions{numbers,sort&compress}

\begin{document}

\begin{frontmatter}

\title{                                                              
Reorientation-effect measurement of the first 2$^+$ state in $^{12}$C:\\ 
confirmation of oblate deformation}                        

\author[a,b]{M. Kumar Raju} 
\author[a]{J. N. Orce\corref{mycorrespondingauthor}}
\cortext[mycorrespondingauthor]{Corresponding author}
\ead{jnorce@uwc.ac.za}
\author[c]{ P. Navr\'atil }
\author[c]{G. C. Ball}
\author[d]{T. E. Drake}
\author[a,b]{S. Triambak}
\author[c]{G. Hackman}
\author[c]{C. J. Pearson}
\author[a]{\\K. J. Abrahams}
\author[a]{E. H. Akakpo}
\author[c]{H. Al Falou}
\author[c]{R. Churchman}
\ead{Deceased.}
\author[e]{D. S. Cross}
\author[c]{M. K. Djongolov}
\author[a]{N. Erasmus}
\author[f]{\\P. Finlay}
\author[c]{A. B. Garnsworthy}
\author[f]{P. E. Garrett}
\author[g]{D. G. Jenkins}
\author[c]{R. Kshetri}
\author[f]{K. G. Leach}
\author[a]{S. Masango}
\author[a]{\\D. L. Mavela}
\author[a]{C. V. Mehl}
\author[a]{M. J. Mokgolobotho}
\author[a]{C. Ngwetsheni}
\author[a]{G. G. O'Neill} 
\author[f]{E. T. Rand}
\author[c]{\\S. K. L. Sjue}
\author[f]{C. S. Sumithrarachchi}
\author[f]{C. E. Svensson}
\author[c]{E. R. Tardiff}
\author[c]{S. J. Williams}
\author[c]{J. Wong}

\address[a]{Department of Physics \& Astronomy, University of the Western Cape, Bellville-7535, South Africa }
\address[b]{iThemba LABS, National Research Foundation, PO Box 722, 7129 Somerset West, South Africa}
\address[c]{TRIUMF, 4004 Wesbrook Mall, Vancouver, British Columbia V6T 2A3, Canada}
\address[d]{Department of Physics, University of Toronto, Toronto, Ontario M5S 1A7, Canada}
\address[e]{Department of Chemistry, Simon Fraser University, British Columbia V5A 1S6, Canada}
\address[f]{Department of Physics, University of Guelph, Guelph, Ontario N1G 2W1, Canada}
\address[g]{Department of Physics, University of York, York YO10 5DD, United Kingdom}

\date{\today}

\begin{abstract}

A  Coulomb-excitation reorientation-effect measurement using the {\small TIGRESS} 
$\gamma-$ray spectrometer at the {\small TRIUMF/ISAC II} facility has permitted the determination of 
the  $\langle 2^+_1\mid\mid \hat{E2} \mid\mid 2^+_1\rangle$ diagonal matrix element in $^{12}$C from 
particle$-\gamma$ coincidence data and state-of-the-art no-core shell model calculations of the nuclear polarizability. 
The nuclear polarizability 
for the ground and first-excited (2$^+_1$) states in $^{12}$C have been calculated 
using chiral {\small NN} {\small N$^4$LO500}  and  {\small{NN+3NF350}} 
interactions,  which show convergence and agreement with photo-absorption cross-section data.  
Predictions show a change in the nuclear polarizability with a substantial increase between the ground state and 
first excited 2$^+_1$ state at 4.439 MeV.   
The  polarizability of the 2$^+_1$ state is introduced into the current and previous 
Coulomb-excitation reorientation-effect analyses of $^{12}$C. 
Spectroscopic quadrupole moments of $Q_{_S}(2_1^+)= +0.053(44)$ eb 
and $Q_{_S}(2_1^+)= +0.08(3)$ eb are determined, respectively, yielding a weighted average of 
$Q_{_S}(2_1^+)= +0.071(25)$ eb, 
in agreement with recent {\it ab initio} calculations.
The present measurement confirms that the 2$^+_1$ state of $^{12}$C is oblate and  emphasizes the 
important role played by the nuclear polarizability in Coulomb-excitation studies of light nuclei.

\end{abstract}

\begin{keyword}
Reorientation effect \sep Spectroscopic quadrupole moment \sep {\it ab initio} calculations 
\sep nuclear polarizability

\end{keyword}
\end{frontmatter}

Electric quadrupole  matrix elements are  key quantities in probing the collective structure of nuclei as 
they are a sensitive and direct measure of the quadrupole deformation. 
The precise determination of these matrix elements reveals the nuclear collectivity caused by the coherent 
motion of nucleons, and the associated nuclear wavefunctions. 
Modern nuclear theory is providing refined calculations of electric quadrupole  matrix elements and related properties 
in light nuclei. Of particular interest is the testing-ground nucleus 
$^{12}$C, as this is computationally accessible to most modern theoretical 
approaches, 
including \emph{ab initio}~\cite{epelbaum2,epelbaum,dreyfuss,kravvaris,peter,forseen,calci,steven,peter2} 
and cluster calculations~\cite{carboncito,neff,FMD,a1,a2,a3}.  
Cluster calculations in $^{12}$C suggest that the admixture of $\alpha$-cluster wavefunctions  may have a pronounced effect on 
the shape of mean-field states at lower energies.
Considerable $\alpha$-cluster triangle admixtures of 52\% and 67\% for the ground and 2$^+_1$ states, respectively,
are predicted by fermionic molecular dynamics ({\small FMD}) calculations~\cite{FMD}, 
whereas a  mean-field contribution 
of 15$\%$ is predicted for the 0$^+_2$ Hoyle state~\cite{Hoyle};
the state crucial to fusion of three $\alpha$ particles in the core of massive stars. 
Moreover, cluster models predict a combination of triangular oblate shapes for the ground state and first 2$^+_1$ excitation 
in $^{12}$C~\cite{neff,carboncito,FMD}. 
Mean-field calculations using a relativistic energy-density functional also show a 
cluster-like structure for the ground state of $^{20}$Ne~\cite{khan}.   


Experimentally, this strong mixing between the 0$^+_2$ Hoyle and 0$^+_1$ ground states 
is supported by the largest known electric monopole transition 
strength, $10^3 \times \rho^2(E0) = 500(81)$, determined from electron scattering measurements~\cite{kibedi}, 
which corresponds to about a 30\% increase in the mean squared charge radius for the  Hoyle state. 


The spectroscopic quadrupole moment, $Q_{_S}(J)$, provides a measure of the extent to which the 
nuclear charge distribution in the laboratory frame  acquires an ellipsoidal shape~\cite{deBoerEichler,hausser0}, 
and can be determined for states with angular momentum $J\neq 0,\frac{1}{2}$~\cite{deShalit}. 
For the 2$^+_1$ state, assuming an ideal rotor, $Q_{_S}(2^+_1)$ is related to the intrinsic quadrupole moment, $Q_{_0}$, in the body-fixed frame 
by $Q_{_S}(2^+_1) = -\frac{2}{7} ~Q_{_0}$~\cite{deShalit}.
Most theoretical approaches predict a very similar $Q_{_S}(2^+_1)\approx+0.06$ eb 
for $^{12}$C~\cite{epelbaum2,forseen,calci,barker,abgrall,bassel},  
which supports a substantial oblate 
deformation.
Recent 
\emph{ab initio}
calculations~\cite{epelbaum2,forseen,calci} provide theoretical uncertainties in their results which makes for more meaningful comparison 
with experiment. Among those worth noting are no-core shell model (NCSM) calculations of $Q_{_S}(2^+_1)$ values with 
unprecedented high precision~\cite{dreyfuss,forseen,calci}.

The reorientation effect~\cite{deBoerEichler,hausser0,alder} ({\small RE}) 
in Coulomb excitation measurements at  energies well below the Coulomb barrier -- so-called safe Coulomb excitation -- 
provides a powerful spectroscopic probe
for extracting $\langle 2^+_1\mid\mid \hat{E2} \mid\mid 2^+_1\rangle$ diagonal matrix elements, 
which can be directly related to the $Q_{_S}(2_1^+)$ value as $Q_{_S}(2_1^+) =0.75793~\langle 2^+_1\mid\mid \hat{E2} \mid\mid 2^+_1\rangle$~\cite{alder}.

The only {\small RE} measurement of the 2$^+_1$ state at 4.439 MeV in $^{12}$C was performed at safe energies
by Vermeer {\it et al.}~\cite{vermeer} through 
a 
measurement of inelastically scattered $^{12}$C ions by a $^{208}$Pb target. 
The scattered $^{12}$C ions were momentum analyzed using a 
magnetic spectrometer and detected at the focal plane 
using a position sensitive multi-wire proportional counter placed at a scattering angle in the 
laboratory frame of $\theta=90^\circ$. 
A value of  $Q_{_S}(2_1^+) = +0.06(3)$ eb was determined using 
the nominal nuclear polarizability parameter $\kappa(g.s.)=1$ 
determined for the ground state of heavier nuclei~\cite{levinger}. 
This parameter represents the ratio of the observed isovector giant-dipole-resonance ({\small GDR}) 
effect to that predicted by the hydrodynamic model~\cite{levinger}, and is a pivotal ingredient in the {\small RE} 
analysis of light nuclei, where $\kappa >1$ values are generally observed~\cite{levinger,orce,vermeer,10be,hausser2,bazhanov,7Li_Vermeer,10B_vermeer,17O}. 


In this work, we perform a safe Coulomb-excitation {\small RE} study of the high-lying 2$^+_1$ state  in $^{12}$C 
using  particle$-\gamma$ coincidence measurements and state-of-the-art {\small NCSM} calculations of the 
nuclear polarizabilities 
$\kappa(g.s.)$ and $\kappa(2^+_1)$. Although there seems to be a good agreement between previous theoretical and 
experimental values of $Q_{_S}(2_1^+)$, the present result 
emphasizes the crucial 
importance of determining $\kappa$ in Coulomb-excitation studies of loosely-bound light nuclei. 
The main advantage of the particle-$\gamma$ coincidence technique lies in the absence of target contaminants 
in the Doppler-corrected $\gamma$-ray spectrum. 

For this  measurement, a beam of $^{12}$C$^{3+}$ ions, delivered to the {\small TRIUMF/ISAC II} facility~\cite{isac} at 4.975 AMeV, 
has been used to populate the 2$^+_1$ state at 4.439 MeV in $^{12}$C through Coulomb excitation. 
The beam energy was chosen in conformity with Spear's prescription of a minimum separation between nuclear surfaces of 
$S(\vartheta)_{min}\gtrapprox 6.5$ fm~\cite{spear} to avoid Coulomb-nuclear interferences, 
where $\vartheta$ is the scattering angle in the center-of-mass frame. 
An average intensity of  $\approx5\times10^8$ particles/s was delivered to the {\small TIGRESS} array~\cite{tigress} 
over approximately three days, 
and impinged on a 3 mg/cm$^2$ thick $^{194}$Pt target (96.45\% enriched). 
The online data were collected in event-by-event mode 
using a high-speed digital data acquisition system with 100 MHz {\small TIG-10}  digital electronics modules. 

The $\gamma$ rays de-exciting  the beam and target nuclei were detected using 
eight {\small TIGRESS} HPGe clover detectors positioned at 14.5 cm from the target, 
and covering around 15$\%$ of 4$\pi$. 
The scattered ions and recoiling particles were detected in a 
double-sided, CD-type silicon detector ({\small S2} type from Micron Semiconductors~\cite{cat}), 
which was mounted 19.4 mm downstream. 
The experimental set up is very similar to the one given 
in Ref.~\cite{10be} apart from the use of an {\small S2} detector, 
which is segmented into 48  rings 
%
and 16 azimuthal sectors on the ohmic side, 
and has the 12 outermost rings incomplete; 
hence, it does not present full azimuthal or $\phi$ symmetry. 
The scattered beam was fully stopped in the 500-$\mu$m thick {\small S2} detector. 
Additional experimental details will be presented in a separate manuscript.

\begin{figure}[!ht]
\begin{center}
\includegraphics[width=7cm,height=6.cm,angle=-0]{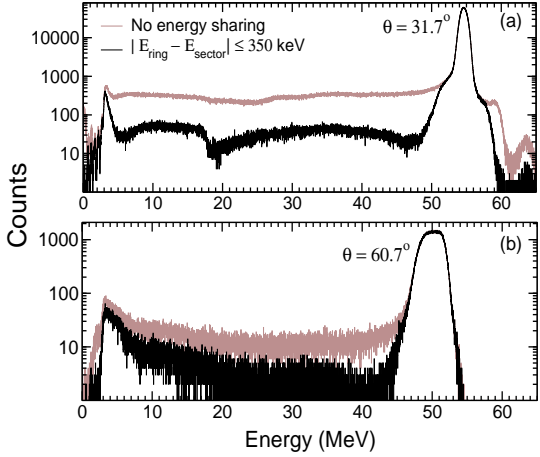}
\caption{Typical particle-energy spectra for the rings at average $\theta$ angles of (a) 31.7$^\circ$ and 
(b) 60.7$^\circ$ obtained 
with (black) and without (light brown) an energy-sharing condition, see text for details.}
\label{particle}
\end{center}
\end{figure}

The energy calibration and relative photo-peak efficiency $\varepsilon$ of the {\small TIGRESS} detectors  were 
determined  using 
standard radioactive $^{152}$Eu and $^{56}$Co sources.  The calibration of all the silicon strips  
was done using a triple $\alpha$ source containing $^{239}$Pu, $^{241}$Am   
and $^{244}$Cm,  together with higher-energy calibration points provided by the elastically scattered beam particles 
simulated with {\small GEANT4}~\cite{geant}, both including energy 
losses~\cite{srim} in the $^{194}$Pt target and the 0.58-mg/cm$^2$ thick aluminum coating on the ohmic side 
of the {\small S2} detector facing the scattered $^{12}$C beam. Typical  particle-energy spectra for the 
innermost (a) and outermost (b) rings at average angles of $\theta=$ 31.7$^{\circ}$ and 60.7$^{\circ}$ 
are shown in Fig.~\ref{particle}.

Particle$-\gamma$ coincidence events were selected by employing the condition that 
each hit in a {\small TIGRESS} detector has a  hit in both a ring ($\theta$) and a 
sector ($\phi$) of the {\small S2} detector within a coincidence time window of approximately 195 ns. 
The corresponding $\gamma-$ray spectra were further cleaned  by subtracting random coincidence events 
outside the 195-ns time window.  
An additional energy-sharing condition 
of $\vert E_{ring}-E_{sector} \vert \leq$ 350 keV 
between each ring and sector 
yields a large 
background reduction in the particle spectra
and enables  a 
better selection of the inelastic peaks~\cite{10be}. 
This energy-sharing condition was chosen to achieve the most optimum background reduction while 
conserving the area of the 
4439-keV peak in the $\gamma-$ray spectrum. 
Figure~\ref{particle}  
illustrates the effect  with a large 
background reduction (black), as compared with no energy-sharing condition (brown), 
at low and intermediate energies. 
Finally, inelastic particle gates can be set on each ring particle spectrum to collect 
solely Coulomb-excitation  events in coincidence with the $\gamma$ ray of interest.  

\begin{figure}[!ht]
\begin{center}
\includegraphics[width=7cm,height=4.5cm,angle=-0]{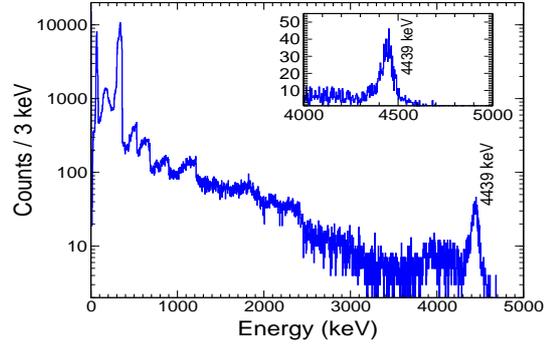}
\caption{Doppler-corrected  $\gamma-$ray energy spectrum generated by employing  particle$-\gamma$, time, energy-sharing 
and inelastic-particle coincidence conditions.
}
\label{gamma}
\end{center}
\end{figure}

Figure \ref{gamma} shows the Doppler-corrected $\gamma-$ray energy spectrum 
obtained from the {\small TIGRESS} array after 
applying  particle and time tagging conditions. 
The  spectrum shows the  328- and 4439-keV $\gamma-$ray transitions depopulating the 2$^+_1$ 
level in $^{194}$Pt and $^{12}$C, respectively. 
Another second-order effect in Coulomb-excitation perturbation theory which may influence, 
particularly for light nuclei~\cite{vermeer,10be,orce}, the determination of both  the sign and magnitude 
of $\langle 2^+_1\mid\mid \hat{E2} \mid\mid 2^+_1\rangle$ is the $E1$ polarizability~\cite{eichler}. 
This involves virtual electric-dipole excitations via the {\small GDR} that polarize the shape of the 
2$^+_1$ state 
through two-step processes of the type $0^+_1 \rightarrow 1^-_{GDR} \rightarrow 2^+_1$.   
In particular, light nuclei present typical values of $\kappa > 1$~\cite{orce}, 
which has a net effect of shifting the measured $Q_{_S}(2_1^+)$ values towards more prolate shapes. 

The polarization potential $V_{pol}$ generated by the $E1$ polarizability is incorporated into  Coulomb-excitation analyses  
by a reduction in  the  quadrupole interaction, $V_{_0}(t)$, which results in an effective potential, $V_{_{eff}}(t)$~\cite{hausser2},
\begin{eqnarray}
 V_{_{eff}}(t) &=&  ~V_{_0}(t)~ \left(~1-V_{pol}(t)\right) \label{eq:pol} \\ 
 &=&  ~V_{_0}(t) \left(1-z\frac{a}{r(t)}\right)\nonumber,
\end{eqnarray}
where 
$a$ is the half-distance of closest approach in a head-on collision and 
$r(t)$ the magnitude of the projectile-target position vector. For the case of 
projectile excitation, 
$z$ is  given by~\cite{alder},
\begin{equation}
z= \frac{10Z_t~\alpha}{3Z_pR^2a} \approx 0.0039~\kappa~\frac{T_{_p} A_{_p}}{Z_{_p}^2(1+A_{_p}/A_{_t})},
\label{eq:0038}
\end{equation}
with  $R=1.2A^{1/3}$ fm being the nuclear radius, 
$T_{_p}$ the kinetic energy (in MeV) in the laboratory frame, 
$\alpha=\frac{\hbar c}{2\pi^2}\sigma_{_{-2}}$ 
the nuclear polarizability, where $\alpha=2P_{_0}$ as defined by Alder and Winther~\cite{alder}, 
and $\kappa$ the polarizability parameter.
The {\small $(-2)$} moment of the total 
photo-absorption cross section, $\sigma_{_{-2}}$,  and $\kappa$ are related by~\cite{orce},  
\begin{equation}
\sigma_{_{-2}}=2.4\kappa~A^{5/3} ~\mu\mbox{b/MeV}. 
\label{eq:mine}
\end{equation}
The value of $\kappa$ can accordingly be modified in modern Coulomb-excitation codes such as {\small GOSIA}~\cite{gosia}. 
For light nuclei, values of $\kappa\ >1$ have been determined by Coulomb-excitation 
measurements for a few favorable cases where $Q_{_S}(J)=0$~\cite{hausser2,7Li_Vermeer,17O}, i.e., for $J=1/2$ excited states, 
and shell-model calculations~\cite{barker,barker2}. 
For the case of arbitrary spins, H\"ausser and collaborators  developed an expression for $\kappa$ 
in terms of $E1$ and $E2$ matrix elements~\cite{hausser2}, $\kappa=\frac{X}{X_{_0}}$, where  
$X_{_0}=0.0004\frac{A}{Z}$~eMeV$^{-1}$ 
and $X$ is given by,
\begin{eqnarray}
X &=& 
\frac{\sum_{n} W(11J_iJ_f, 2J_n) ~\frac{\langle i\mid\mid \hat{E1} \mid\mid n\rangle \langle n\mid\mid \hat{E1} \mid\mid f\rangle}{E_n - E_i}}{\langle i\mid\mid \hat{E2} \mid\mid f\rangle}, 
\label{eq:se1}
\end{eqnarray}
\noindent where the sum extends over all intermediate states $|n\rangle$ connecting the  
initial $|i\rangle$ and final $|f\rangle$  states with $E1$ transitions 
and  $W(11J_iJ_f, 2J_n)$  is the Racah W-coefficient~\cite{racah} with  $J_i=0$, $J_f=2$ and $J_n=1$ for the case at hand. 
It is important to note here
that the product of two $\hat{E1}$ operators yields an $\hat{E2}$ operator; hence, some of the isoscalar
giant quadrupole resonance strength may appear in the sum given in Eq.~\ref{eq:se1}.


\begin{figure}[!ht]
\begin{center}
\includegraphics[width=6.cm,height=4.5cm,angle=-0]{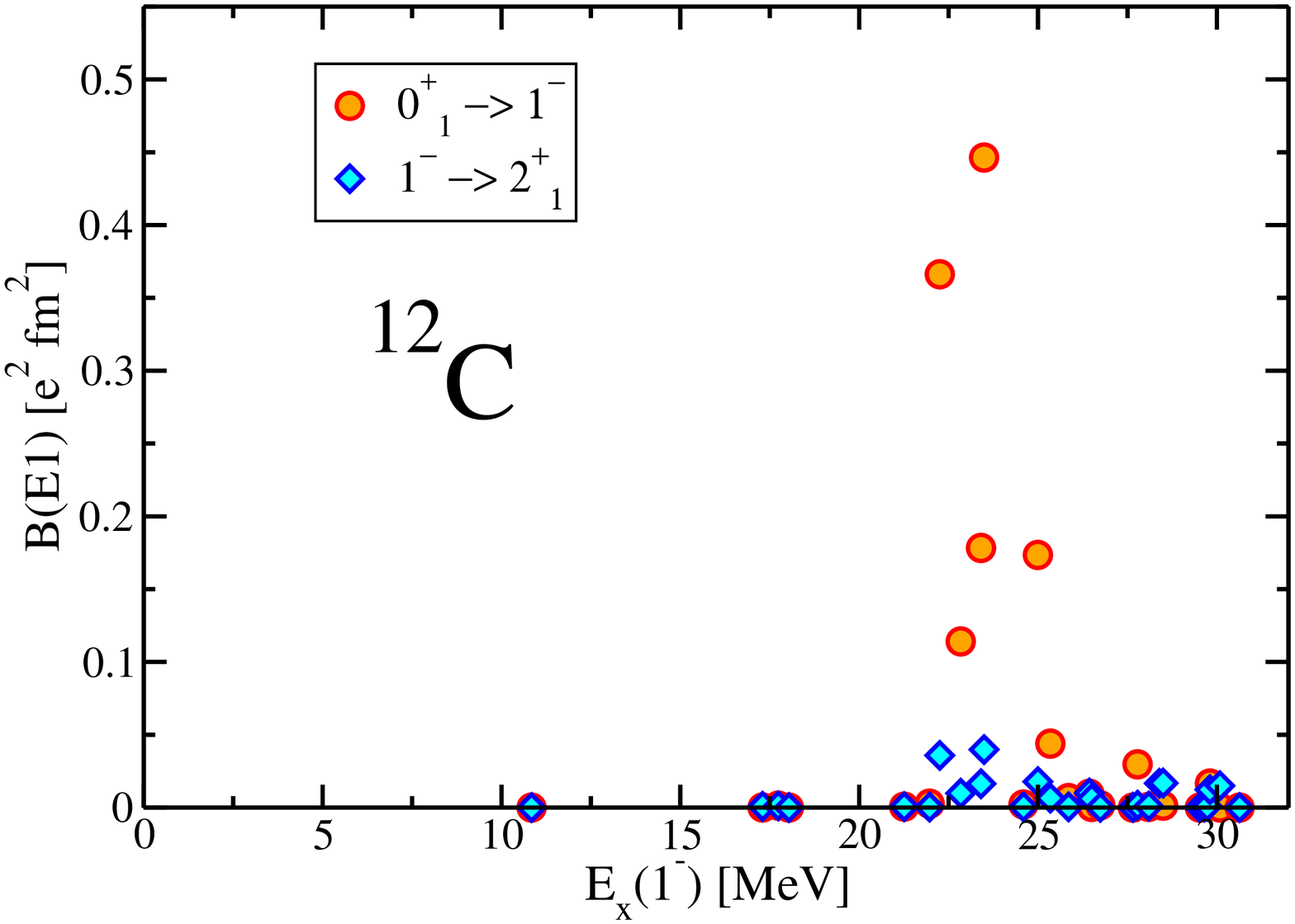}
\caption{$B(E1)$ strengths calculated with the {\small NCSM} using the chiral $NN+3NF350$  interaction 
for $0^+_1\rightarrow 1^-$ and $1^-\rightarrow2^+_1$ transitions.}
\label{BE1}
\end{center}
\end{figure}

In the present work, {\small NCSM} calculations have been performed  to estimate 
$\kappa$ for the ground and 2$_1^+$ states in $^{12}$C. Previous {\small SM} calculations 
of $\kappa(2_1^+)=0.77$ presumed that all the $E1$ strength from 
the ground state was concentrated at the {\small GDR}  energy~\cite{barker}. 
Our {\small NCSM}
calculations used the chiral {\small $NN+3NF350$}  interaction~\cite{entem1,navra1,navra2}, 
including the N$^3$LO $NN$ interaction~\cite{entem1} and the local N$^2$LO $3N$ interaction~\cite{navra1} 
with the cutoff of 350 MeV~\cite{navra2}, 
and considered model spaces with basis sizes of $N_{max}=4$ for the natural and 
$N_{max}=5$ for the unnatural parity states. From Eq.~\ref{eq:se1}, which included the $E1$ matrix elements from all 
the transitions connecting 28 1$^-$ states up to 30 MeV, values of $\kappa(g.s.)=1.6(2)$ and $\kappa(2_1^+) = 2.2(2)$   
are predicted.  As shown in Fig. \ref{BE1},  the $E1$ strength 
is concentrated at an energy of about 24 MeV -- the centroid energy of the {\small GDR}~\cite{atlas}. 
The lowest calculated 1$^-$ state energy was set to the lowest found 1$^-_1$ state at 10.84 MeV. 
In order to study convergence and determine uncertainties, predictions with the {\small $NN+3NF350$} 
interaction have been validated by  
additional {\small NCSM} calculations using the  {\small $NN$} 
{\small N$^4$LO500} interaction~\cite{navra3,entem2} {\small SRG} evolved~\cite{navra4} to 2.4 fm$^{-1}$ 
at the same $N_{max}=4/5$ space and at a smaller $N_{max}=2/3$ space at varied harmonic-oscillator frequencies, 
as well as at a larger $N_{max}=6$ space for natural parity and $N_{max}=7$ for unnatural parity states, 
which included 22 1$^-$ states up to 30 MeV. The latest calculations yield similar results of $\kappa(g.s.)=1.5(2)$ and 
$\kappa(2_1^+) = 2.1(2)$.  In general, to improve on the present {\small NCSM} description, 
one should include RGM-like cluster states with explicit $\alpha$ particles, e.g., $^8$Be+$\alpha$ and couple them with the currently 
used {\small NCSM} basis. Such approach called {\small NCSM} with continuum is now under development.  
However, the good stability of all the 1$^-$ states with $N_{max}$ demonstrates that our
expansion is adequate.


The well-known total photo-absorption cross section measured for the ground state of $^{12}$C can be used to 
benchmark our {\small NCSM} calculations. 
A value of $\sigma_{_{-2}}=244$ $\mu$b/MeV  in the 1985 evaluation by Fuller~\cite{fuller},  
yields $\kappa(g.s)=1.6$ using Eq.~\ref{eq:mine}, in excellent 
agreement with our {\small NCSM} polarizability calculations for the ground state. 
The consistency of our calculations further supports the 
value of $\kappa(2_1^+) = 2.2(2)$ implemented in our {\small GOSIA} analysis~\cite{gosia} throughout this work. 

\begin{figure}[!h]
\begin{center}
\includegraphics[width=6.5cm,height=6.cm,angle=-0]{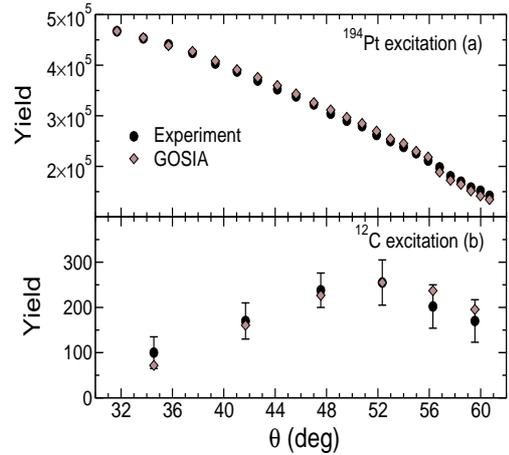}
\caption{Heavy-ion angular distributions showing experimental
and calculated $\gamma-$ray yield as a function of laboratory scattering angle, $\theta$, for the de-excitation of the 2$_1^+$ states  
in $^{12}$C (bottom) and $^{194}$Pt (top), see text for details.}
\label{ang}
\end{center}
\end{figure}

\begin{figure}[!ht]
\begin{center}
\includegraphics[width=6.8cm,height=5.cm,angle=-0]{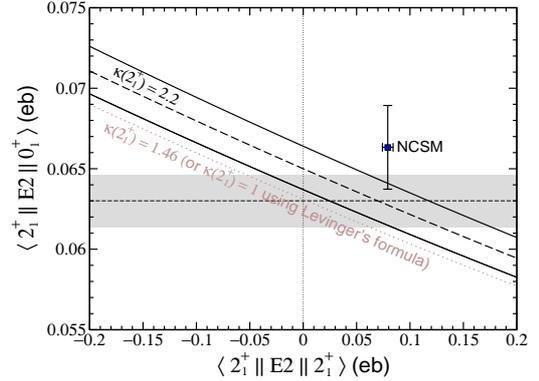}
\caption{Variation of $\langle 2^+_1\mid\mid \hat{E2} \mid\mid 0^+_1\rangle$ as 
a function of $\langle 2^+_1\mid\mid \hat{E2} \mid\mid 2^+_1\rangle$  in $^{12}$C for $k(2^+_1)=2.2$. 
The  horizontal band represents the 1-$\sigma$ boundary for 
$\langle 2^+_1\mid\mid \hat{E2} \mid\mid 0^+_1\rangle=0.0630(16)$~\cite{be2}. 
For comparison, the square data point shows the result from high-precision {\small NCSM} calculations, 
$Q_{_S}(2_1^+)= +0.060(4)$ eb and $B(E2; 2^+_1 \rightarrow 0^+_1)=8.8(7)$ e$^2$fm$^4$~\cite{forseen}.}
\label{coulex}
\end{center}
\end{figure}

The integrated $\gamma-$ray yields for the $2_1^+\rightarrow 0_1^+$ transitions 
in  $^{12}$C and  $^{194}$Pt have been calculated using the semi-classical 
coupled-channel Coulomb-excitation least-squares code {\small GOSIA}~\cite{gosia}. 
The semi-classical approximation  is  confirmed from Rutherford scattering cross sections 
and the calculated Sommerfeld parameter, $\eta=\frac{a}{\lambdabar}=31\gg1$, where 
$\lambdabar$ is the de Broglie  wavelength. 
Calculations consider the known spectroscopic information such as level lifetimes, 
branching ratios and matrix elements,  
kinematics, detector geometry and beam energy losses. 
The effect of higher-lying states in $^{12}$C has been estimated using {\small GOSIA} and considered negligible ($< 0.1\%$). 
Figure \ref{ang} shows  the experimental and theoretical heavy-ion angular distributions 
from the eight clover yields for the  $2^+_1\rightarrow 0^+_1$ transitions in  $^{194}$Pt (a) and $^{12}$C (b). 
Predictions of the cross sections for populating states in $^{12}$C were  
calculated at fixed values of $\langle 2^+_1\mid\mid \hat{E2} \mid\mid 0^+_1\rangle  = 0.0630$ eb~\cite{be2}, 
$\langle 2^+_1\mid\mid \hat{E2} \mid\mid 2^+_1\rangle = +0.070$ eb and $\kappa=2.2$, 
the intersection point of the centroid of the two bands in Fig.~\ref{coulex}, 
and normalized to the experimental yields with a common normalization factor. 
The shape of the angular distributions predicted by {\small GOSIA} for both $^{194}$Pt  and $^{12}$C are in 
good agreement with experiment. 

The normalization procedure used in Ref.~\cite{10be} was applied to determine $\langle 2^+_1\mid\mid \hat{E2} \mid\mid 2^+_1\rangle$, 
where Coulomb-excitation curves are determined in the 
$\langle 2^+_1\mid\mid \hat{E2} \mid\mid 2^+_1\rangle-\langle 2^+_1\mid\mid \hat{E2} \mid\mid 0^+_1\rangle$ plane 
by fixing $\langle 2^+_1\mid\mid \hat{E2} \mid\mid 2^+_1\rangle$ in steps of 0.01 eb,  
and varying $\langle 2^+_1\mid\mid \hat{E2} \mid\mid 0^+_1\rangle$ 
until converging with  the  experimental intensity ratio between target and projectile, $I_{_\gamma}^T/I_{_\gamma}^P$, 
given by, 
\begin{equation}\label{normalisation}
\frac{\sigma_{_{E2}}^T  W(\vartheta)^T}{\sigma_{_{E2}}^P  W(\vartheta)^P} = 
{\small 1.037}~\frac{N_{_\gamma}^T}{N_{_\gamma}^P}\frac{\varepsilon_{_\gamma}^P}{\varepsilon_{_\gamma}^T} = \frac{I_{_\gamma}^T}{I_{_\gamma}^P},
\end{equation}
\noindent where $W(\vartheta)$  represents the integrated angular distribution of the de-excited $\gamma$ rays 
in coincidence with the inelastic scattered particles~\cite{alder2} and 
the factor {\small 1.037} accounts for the 96.45\% enrichment of the $^{194}$Pt target chosen for normalization. 
The normalization of the $\gamma$-ray yield in 
$^{12}$C to the well-known matrix elements 
in the target nucleus, $^{194}$Pt, minimizes systematic effects such as dead time and pile-up rejection. 
Absolute efficiencies of $\varepsilon_{_\gamma}^P = 0.0162(5)$ and $\varepsilon_{_\gamma}^T = 0.0784(8)$, and 
total counts of $N_{_\gamma}^P=1150(40)$ and $N_{_\gamma}^T=7021190(2650)$  for the 4439- and 328-keV $\gamma-$ray transitions, 
respectively, yield $I_{\gamma}^T/I_{\gamma}^P=1308(62)$. 
The quoted error on this measurement arises from the uncertainties of $N_{_\gamma}^P$ (3.5\%) 
and $\varepsilon_{_\gamma}^P$ (3.1\%). Other contributions are less significant and include the  
$\phi$ asymmetry of the 
{\small TIGRESS} detectors ($<0.5\%$)~\cite{21na}.


The resulting Coulomb-excitation diagonal band is shown in Fig.~\ref{coulex}, 
where the black dashed line is the central value and the two black solid lines correspond 
to the $1\sigma$ loci limits. 
The horizontal band represents $\langle 2^+_1\mid\mid \hat{E2} \mid\mid 0^+_1\rangle=0.0630(16)$ eb~\cite{be2}.
Assuming $\kappa(2_1^+) = 2.2$, a positive value of 
$\langle 2^+_1\mid\mid \hat{E2} \mid\mid 2^+_1\rangle=+0.070(71)$ eb  is obtained from 
the intersection of the two bands, corresponding to $Q_{_S}(2^+_1)=+0.053(53)$ eb.  
The error of $\langle 2^+_1\mid\mid \hat{E2} \mid\mid 2^+_1\rangle$ was determined 
from the overlap region 
between the two bands assuming central values for the 
$\langle 2^+_1\mid\mid \hat{E2} \mid\mid 0^+_1\rangle$ band, $\pm0.045$ eb, 
and the Coulomb-excitation diagonal curve, $\pm0.055$ eb, added in quadrature. 
The uncertainty of $\kappa(2_1^+)$, $\pm0.01$ eb, yield final values 
of $\langle 2^+_1\mid\mid \hat{E2} \mid\mid 2^+_1\rangle=+0.070(72)$ eb and $Q_{_S}(2^+_1)=+0.053(54)$ eb.  
Moreover, if one uses $\kappa(2^+_1)=1$ assuming Levinger's formula~\cite{levinger}, 
$\sigma_{_{-2}}=3.5\kappa A^{5/3}$ $\mu$b/MeV (which corresponds to $\kappa(2^+_1)=1.46$ 
using Eq.~\ref{eq:mine}), 
our data yields  $Q_{_S}(2_1^+)=+0.003(54)$ eb, as shown by the dotted brown line in Fig.~\ref{coulex}; 
a value  consistent with a spherical shape.



A more precise determination of the statistical uncertainty of $\langle 2^+_1\mid\mid \hat{E2} \mid\mid 2^+_1\rangle$ 
has been done by employing the error minimization procedure in {\small GOSIA}~\cite{coulex-analysis},  
considering $\langle 2^+_1\mid\mid \hat{E2} \mid\mid 0^+_1\rangle = 0.0630(16)$ eb~\cite{be2}
and $\langle 2^+_1\mid\mid \hat{E2} \mid\mid 2^+_1\rangle = 0.070(72)$ eb 
as initial inputs along with available matrix elements of higher-lying states. 
Using the six experimental 
yields given in Fig.~\ref{ang}, 
the error minimization  carried out in a two-step process, by calculating the uncorrelated and correlated 
errors, yields a final error  of $\Delta\langle2^+_1\mid\mid\hat{E2} \mid\mid2^+_1\rangle = 0.058$ eb,  
which includes the error of 
$\kappa(2^+_1)$, $\pm0.01$ eb. 
A final value of $Q_{_S}(2_1^+)= +0.053(44)$ eb is determined, which  accounts for an additional 5\% systematic uncertainty 
in the {\small GOSIA} calculation.  
The main source of systematic uncertainty is attributed to quantal effects, which are inversely proportional 
to $\eta$~\cite{alder,smilansky,alder3,kavka}, and could affect the 
validity of the semi-classical approximation. For $\eta \approx 31$, quantal effects may add an 
uncertainty of $\leq3.5$\% to the present determination of the $Q_{_S}(2_1^+)$ value. 
If one takes the data from Vermeer {\it et al.}~\cite{vermeer} and assumes $\kappa(2^+_1)=2.2$ 
and $\langle 2^+_1\mid\mid \hat{E2} \mid\mid 0^+_1\rangle  = 0.0630(16)$ eb, 
a  potentially more pronounced value of $Q_{_S}(2_1^+)\approx+0.08(3)$ eb is determined, 
in agreement with the present work.  
The weighted average of the current and previous work yields a final value of $Q_{_S}(2_1^+)= +0.071(25)$ eb.

The weighted $Q_{_S}(2_1^+)$ value can be compared with 
state-of-the-art {\it ab initio} calculations. The high-precision \emph{\small NCSM} calculation using the {\small CDB2k NN} potential 
is given in Fig.~\ref{coulex} by the square data point~\cite{forseen}, $Q_{_S}(2_1^+)= +0.060(4)$ eb and 
$B(E2; 2^+_1 \rightarrow 0^+_1)=8.8(7)$ e$^2$fm$^4$. Similar values of  $Q_{_S}(2_1^+)=+0.0591(25)$ eb 
and $Q_{_S}(2_1^+)=+0.059(1)$ eb  are calculated, respectively, using chiral {\small NN+3N} interactions~\cite{calci} and the 
no-core symplectic model~\cite{dreyfuss}. Calculations are in agreement with the weighted average presented in 
this work.

Unfortunately, the model space that can currently be reached with the {\small NCSM} does not allow a 
calculation of the nuclear polarizability for the 2$^+_2$ state built on the Hoyle state.  
One could, however, speculate that if $\kappa$ further increases with excitation energy, as the nucleus becomes 
more loosely bound, a more 
pronounced prolate shape might be expected for the shape of the Hoyle state rotational band. 
This is in concordance with the  
prolate bent-arm configuration -- with $Q_{_S}(2^+_2)=-0.07(2)$ eb -- predicted  by {\it ab initio} calculations 
using chiral perturbation theory on a lattice~\cite{epelbaum}, 
and, although with an extremely large prolate deformation, the $Q_{_S}(2^+_2)=-0.21(1)$ eb value 
predicted by the no-core symplectic model~\cite{dreyfuss}. 
Such an enhanced polarizability might explain the sudden change in the shape of the 
Hoyle state, which seems to be in disagreement with early models of cluster formation such as that of Morinaga, where 
$\alpha$-cluster structures gradually emerge with increasing excitation energy and are fully realized 
at the $\alpha$ threshold~\cite{morinaga,ikeda}.

In conclusion, the Coulomb-excitation analysis performed in this work using the {\small TIGRESS} array 
and the new value of $\kappa(2^+_1)$ calculated with the {\small NCSM} have permitted the determination of 
the  $\langle 2^+_1\mid\mid \hat{E2} \mid\mid 2^+_1\rangle$ diagonal matrix element in $^{12}$C from 
particle$-\gamma$ coincidence data. The present work confirms  
an oblate deformation for the 2$^+_1$ state in $^{12}$C, in agreement with recent {\it ab initio} and cluster-model 
calculations. 

Finally, 
it is important to emphasize   
that {\small  NCSM} calculations show that the polarizability parameter for excited states 
can be very different from the ground state value determined from total photo-absorption 
cross-section data.
This unanticipated change of the nuclear polarizability from the ground state to the first excitation 
in $^{12}$C may not only affect Coulomb-excitation analyses of light nuclei, but in general, 
as nuclei become less bound, Coulomb-excitation studies of states at high excitation energies 
(e.g., superdeformed bands found in nuclei with spherical ground states). 
This possibility clearly needs further investigations.





\section*{Acknowledgements} 
										
We thank the accelerator group  at TRIUMF for their support during the experiment. 
This work has been partially supported by the South African 
National Research Foundation (NRF) under Grants 93500 
and the Natural Sciences and Engineering Research Council of Canada. 
{\small TRIUMF} receives federal funding via a contribution agreement through the National Research Council of Canada. 
JNO thanks J. L. Wood for pointing to the $\hat{E2}$ operator relationship in Eq.~\ref{eq:se1} and the late D. H. Wilkinson 
for  fruitful discussions. Computing support came from an INCITE Award on the Titan supercomputer of the Oak Ridge 
Leadership Computing Facility (OLCF) at ORNL, and from Calcul Quebec and Compute Canada. 



\end{document}